\documentclass[12pt]{article}
\def\be{\begin{equation}}
\def\ee{\end{equation}}
\def\bea{\begin{eqnarray}}
\def\eea{\end{eqnarray}}

\def\aaa{{\cal A}}
\def\bbb{{\cal B}}
\def\p{\partial}
\def\eqq{\stackrel{\sigma}{=}}
\def\sup{\sigma}
\def\suup{\sigma}
\def\mmm{{\cal V}}%(halves with boundary)
\def\mm2{{\cal V}}%(complete ones)
\def\d{d}
\def\g{g}
\newcommand{\bm}[1]{\mbox{\boldmath $#1$}}

\def\CA{\check{A}}
\def\CB{\check{B}}
\def\CC{\check{C}}
\def\CD{\check{D}}
\def\CE{\check{E}}
\def\BA{\hat{A}}
\def\BB{\hat{B}}
\def\BC{\hat{C}}
\def\BD{\hat{D}}
\def\BE{\hat{E}}
\def\tpp{\tilde{\varphi}}
\def\tz{\tilde{z}}

\def\Partp{\frac{\partial}{\partial \tilde{\varphi}}}
\def\Partz{\frac{\partial}{\partial \tilde{z}}}
\def\Parrho{\frac{\partial}{\partial \rho}}
\def\Partt{\frac{\partial}{\partial T}}

\def\Parsz{\frac{\partial}{\partial \zeta}}
\def\Parsl{\frac{\partial}{\partial \lambda}}

\def\Journal#1#2#3#4#5#6{{#1} {\bf #2} (#5) #3}
\def\CQG{\em Class. Quantum Grav.}

\def\GRG{\em Gen. Rel. Grav.}

\def\PR{\em Phys. Rev.}

\def\JMP{\em J. Math. Phys.}

\def\CMP{\em Commun. Math. Phys.}

\def\MPL{\em Mod. Phys. Lett.}

\def\delt{\triangle}
\def\deltt{\hat\triangle}
\def\fin{\hfill \rule{2.5mm}{2.5mm}\\ \vspace{0mm}}
\def\finn{\hfill \rule{2.5mm}{2.5mm}}

\newtheorem{lemma}{Lemma}[section]
\newtheorem{prop}{Proposition}[section]

\newtheorem{coro}{Corollary}[section]
\newtheorem{theorem}{Theorem}[section]
\def\bean{\begin{eqnarray*}}
\def\eean{\end{eqnarray*}}

\begin{document}
%_____________________________________________________________________________
\title{Generalisation of the Einstein-Straus model to anisotropic settings}
%\title{The Einstein-Straus problem in anisotropic settings}
%_____________________________________________________________________________
\author{Filipe C. Mena, Reza Tavakol \& Ra\"ul Vera\\\\
%_____________________________________________________________________________
School of Mathematical Sciences,
Queen Mary, University of London\\ 
Mile End Road                   
London E1 4NS, UK}
%_____________________________________________________________________________
\date{April 2002}
%_____________________________________________________________________________
\maketitle
%_____________________________________________________________________________
\begin{abstract}
%_____________________________________________________________________________
We study the possibility of generalising the Einstein--Straus model
to anisotropic settings, by considering the
matching of
locally cylindrically symmetric static
regions to the set of $G_4$ on $S_3$ locally rotationally symmetric
(LRS) spacetimes.
We show that such matchings preserving the symmetry are only 
possible for a restricted subset of the LRS
models
in which there is no evolution in one spacelike direction.
These results are applied to
spatially homogeneous (Bianchi) exteriors
where the static part represents a finite bounded
interior region without holes.
We find that it is impossible to embed finite static
strings or other locally cylindrically symmetric static objects
(such as bottle or coin-shaped objects)
in reasonable Bianchi cosmological models,
irrespective of the matter content.
Furthermore, we find that if the exterior spacetime
is assumed to have a
perfect fluid source satisfying the dominant energy condition, then
only a very particular family of LRS stiff fluid solutions
are compatible with this model.

Finally, given the
interior/exterior duality in the matching procedure,
our results have the interesting consequence
that the Oppenheimer-Snyder model of collapse
cannot be generalised to such anisotropic cases.

%_____________________________________________________________________________
\end{abstract}
\newpage
%_____________________________________________________________________________
\section{Introduction}
%_____________________________________________________________________________
An important long standing question in cosmology concerns the way large scale
dynamics of the universe influences the behaviour on smaller scales.
In particular, given the observed large scale expansion of the universe
the question is to what extent does this expansion 
influence the behaviour on astrophysical scales, and 
more precisely what are its effects on e.g. the planetary orbits, galaxies and
clusters of galaxies.
Among the earliest works on this question 
are those by McVittie \cite{McVittie} and Einstein \& Straus
\cite{Einstein-Straus} (see
 \cite{Krasinski} for more historical references).
Historically it was 
McVittie who first found a perfect fluid spherically symmetric 
solution to Einstein's field equations which could be interpreted as describing
a point particle embedded in an
expanding Friedmann-Lema\^{\i}tre-Robertson-Walker (FLRW)
spacetime \cite{McVittie}. This 
interpretation has been questioned
by some authors subsequently
(e.g. Sussman \cite{Sussman88}, Gautreau \cite{Gautreau} and
Nolan \cite{Nolan98}). The
generally accepted ansatz to model the problem is due to
Einstein and Straus \cite{Einstein-Straus}, who proposed a matching between
two spacetimes, 
instead of trying to
use a single solution. They
successfully matched the spherically symmetric vacuum 
Schwarzschild solution
to an expanding {\em dust\/}\footnote{This is
a consequence of the FLRW model
having been matched to a vacuum spacetime (see e.g.
\cite{OPSN}).}
FLRW exterior across a hypersurface preserving the symmetry.
They showed that such a matching was possible
across any comoving 2-sphere, as long as the total mass
contained inside the 2-sphere was equal to the 
Schwarzschild mass contained in it.
In this way they concluded that there was no influence from
the global expansion of the universe on
the vacuum region surrounding the Schwarzschild mass.
Two objections have been raised against this model: 
the first by Krasi\'{n}ski \cite{Krasinski} who suggested 
that the Einstein-Straus model is unstable against radial perturbations, and
the second by Bonnor 
\cite{Bonnor000,Bonnor00} who pointed out that there are severe restrictions on
the scales on which the model 
is applicable and that it is not suitable for studies within our solar
system or even the galaxy.    

Another important observation regarding the result of Einstein \& Straus
is that it involves a number of idealisations, including the 
fact that the universe is assumed to be representable by
a spatially homogeneous and isotropic dust FLRW model. 
The question then arises as to whether this result
is robust with respect to various
plausible generalisations. These could involve changes in the 
symmetry properties of the model as well as the nature of
the interior source-field, which was originally 
taken to be vacuum.

A number of interesting attempts have been made in 
this direction. Among them are
models which keep the
spherical symmetry but generalise the interior source fields
by considering for example
Vaidya (see \cite{Fayos-etal} and references therein\footnote{The
aim of these
studies was, in fact, to generalise to non-dust FLRW models.})
or Lema\^{\i}tre-Tolman spacetimes
(see \cite{Krasinski}
for  references concerning the latter in connection 
with formation of voids).
There have also been attempts concerning the relaxation of 
the spherical symmetry assumption, including generalisations to
locally cylindrically symmetric spacetimes. These include
the example of the embedding of
dust FLRW into a non-static vacuum exterior
across a hypersurface of a constant radius \cite{Cocke}
which, due to the freedom in interpreting the two parts being matched
as interior or exterior \cite{Fayos-etal} (which we shall refer to as
{\it interior/exterior duality}), is 
%mathematically
equivalent to an embedding of a non-static vacuum region into a dust FLRW.
Similarly,
the impossibility of the embedding
of typical cosmic strings (i.e. Minkowski with deficit angle)
as well as some special non-static cylindrically symmetric
vacuo into {\em flat} FLRW was shown in
\cite{Shaver-Lake}.
This problem has been further studied by 
Senovilla \& Vera \cite{Seno-Vera} 
who have shown in full generality
that the embedding of a locally orthogonally transitive (OT)\footnote{
%This just a technical assumption as explained below and for
For most matter contents one is interested in
%(vacuum static cavities, for instance), 
this assumption is actually a consequence
of having an axis of symmetry \cite{Carot-etal}, see also below.}
cylindrically symmetric {\em static} cavity
in an expanding FLRW is not possible,
%((across a non-spacelike hypersurface))
irrespective of the matter content of the cavity\footnote{We
note that non-expanding exteriors can be matched, see \cite{Bonnor-etal98}.}.
This result has been further generalised by Mars
\cite{Mars} to the case of axial symmetry.
Mars \cite{Mars01} has finally been able to prove that in order to embed 
{\em any} static cavity in a FLRW universe then this cavity must be ``almost 
spherically symmetric''; more precisely, the boundary as seen from the
FLRW exterior is required to be a 2-sphere in space. 
Furthermore, for standard interior source fields such as
vacuum, electrovacuum or perfect fluids, the interior itself must be
spherically symmetric, with the boundary  
comoving with the cosmological flow
\cite{Mars,Mars01}. This implies therefore that static 
objects which can be embedded
in FLRW models must be spherical, and as a result 
the Einstein--Straus model is, in this sense, not robust.
 
This result once again raises the question of the possibility of 
embedding general static cavities in more general universe models (which we 
refer to as the {\it Generalised Einstein--Straus problem}).
Also, since realistic cosmological models
cannot be expected to be exactly homogeneous and isotropic,
the question arises as to what happens if these symmetry assumptions 
concerning the exterior metric are further 
relaxed. 

There are two different ways to study departures from FLRW: either
perturbatively (see \cite{Ch} for a perturbed generalisation
of Einstein-Straus with a small rotation)
or using exact solutions. Given that 
a precise formalism for a perturbed matching of
two spacetimes is not fully developed,
we shall proceed in the second way.
A step in this direction was taken by Bonnor \cite{Bonnor00},
who considered the embedding of a Schwarzschild region in an expanding
spherically symmetric inhomogeneous Lema\^{\i}tre-Tolman exterior. 
He found that such matching is possible
in general, and it allows the mass and radius for the
Schwarzschild cavity to be chosen independently of the exterior LT density. 

An interesting question is whether similar results would hold for cases
with non-spherically symmetric interiors.
As a step
in this direction, we shall first of all study the
{\em local} matching between static
OT cylindrically symmetric spacetimes and the class of
locally rotationally symmetric (LRS)
spacetimes admitting a $G_4$ on $S_3$ group
of isometries, which constitute an anisotropic generalisation 
of the FLRW models.
%The matching procedure is local. 
To make the matching global
one expects to have further restrictions.
In the particular case
of an interior that describes a bounded object without holes,
we were able to show that this is in fact the case.
Thus if the exterior is assumed
to be a spatially homogeneous expanding Bianchi spacetime,
then, in order to preserve the symmetry, it has to be locally
rotationally symmetric, admitting a $G_4$ on $S_3$.
We note that the results obtained here for the LRS spacetimes also hold
for models representing static cavities embedded in
Bianchi spacetimes.

Our main result is then that 
no locally OT cylindrically
symmetric static cavities can be embedded into
reasonable evolving anisotropic (Bianchi) spacetimes.
%We shall show that for exteriors spacetimes with perfect-fluid content,
%only stiff fluid equations of state are allowed.

We also note that given the interior/exterior duality in the matching
procedure, all our results apply equally to the case where it is the
interior that is taken to have a spatially homogeneous geometry,
embedded into a locally cylindrically
symmetric static background. This would allow 
our results to be applied to other settings, such as
the study of the generalisation of the Oppenheimer-Snyder \cite{OPSN}
model for collapsing objects, which could be
viewed as the `dual' to the Einstein--Straus model,
in the interior/exterior sense defined here.

The plan of the paper is as follows. In Section \ref{geom}
we review the matching procedure and the definition of
matching preserving the symmetry.
In Section \ref{compact-lrs} we present a
compact form of the line-element
for the $G_4$ on $S_3$ LRS spacetimes
in coordinates adapted to the axial Killing vector field.
This will prove useful in sections \ref{parame}   
and \ref{junction}, where we calculate 
the matching
conditions for the matching preserving the symmetry between a static 
OT cylindrically symmetric spacetime and a LRS homogeneous spacetime.
In Section \ref{sec:lemmas} we study the restrictions 
on the subset of LRS spacetimes 
that are allowed by the matching conditions.
We show that the only perfect fluid solutions
in this subset correspond to a particular family
which has a stiff fluid equation of state.
In Section \ref{axialglobal} we extend our results to the
case of the spatially homogeneous exteriors.
Finally, Section \ref{conclusion} gives our discussions and conclusions.

%_____________________________________________________________________________
\section{Matching procedure}
\label{geom}
%____________________________________________________________________________
In this section we shall briefly recall the 
matching procedure across general hypersurfaces
(see \cite{Mars-Seno} for more details).
As is well known, the matching of two spacetimes
requires two sets of (matching or junction) conditions
at the matching hypersurface.
The first set of these junction conditions will ensure the continuity of the metrics
across the matching hypersurface; while
the second is equivalent to a non-singular
Riemann tensor distribution in order to prevent infinite
discontinuities of matter and curvature across the matching
hypersurface.

More precisely, let $(\mmm^+,g^+)$ and $(\mmm^-,g^-)$ be two $C^3$ space-times
with oriented boundaries $\sup^+$ and $\sup^-$ respectively, such that
$\sup^+$ and
$\sup^-$ are
diffeomorphic. The matched space-time
$(\mmm,g)$ is the disjoint union of $\mmm^\pm$
with the points in $\sup^\pm$ identified
such that
the junction conditions are satisfied
(see \cite{Lich,Israel,Clarke-Dray,Mars-Seno}).
Since $\sup^\pm$ are diffeomorphic, one can view 
these boundaries as
diffeomorphic to a
3-dimensional oriented manifold $\sup$ which can be 
embedded in $\mm2^+$ and $\mm2^-$.
Let $\{\xi^a\}$ $(a=1,2,3)$
and
$\{x^{\pm\alpha}\}$ be coordinate systems on $\sup$ and $\mm2^\pm$,
respectively.
The two boundaries
are given by two $C^3$ maps
\bea
\label{embed}
\Phi^\pm:\; \sup &\longrightarrow& \mm2^\pm\\
\label{eq:embeddings}
              \xi^a &\mapsto& {x^\alpha}^\pm={\Phi^\alpha}^\pm(\xi^a),
\nonumber
\eea
such that
$\sup^\pm = \Phi^\pm(\sup)$ %\subset \mm2^\pm$.
At every point $p\in \sigma$
the natural basis $\{\partial/\partial \xi^a|_p\}$
of the tangent plane $T_p \sup$
is pushed-forward by the rank-3 differential maps $\d\Phi|_p^\pm$ 
into three linearly independent
vectors at $\Phi^\pm (p)$, denoted by $\vec e^{\;\pm}_a|_{\Phi^\pm(p)}$,
defined only in the corresponding hypersurfaces $\sup^\pm$,
as follows
% \footnote{For the sake of simplicity in the notation,
% we have dropped any specification to the point $p\in \sigma$
% to which tangent plane the differential map is applied.}
\begin{equation}
\label{push}
\d \Phi^\pm\left(
%\left.
\frac{\partial}{\partial \xi^a}
%\right|_{\suup}
\right)
=\frac{\partial \Phi^{\pm\mu}}{\partial \xi^a}
\left.\frac{\partial}{\partial x^{\pm\mu}}\right|_{\suup^\pm}
\equiv \vec e^{\;\pm}_a%|_{{}_{\suup^\pm}}
=e^{\pm\mu}_a
\left.\frac{\partial}{\partial x^{\pm\mu}}\right|_{\suup^\pm}.
\end{equation}
%On the other hand, 
Using the pull-backs $\Phi^{\pm*}$,
the metrics $g^\pm$ at any point $\Phi^{\pm}(p)\in\sigma^\pm$
are mapped to the dual space at $p\in\sup$ providing
two symmetric 2-covariant tensors $\bar{g}^+$ and $\bar{g}^-$,
whose components in the natural basis $\{\d \xi^a\}$ are
$
\bar{g}^\pm_{ab}= e^{\pm\mu}_a e^{\pm\nu}_b g_{\mu\nu}|_{{}_{\suup^\pm}}=
(\vec e^{\;\pm}_a\cdot \vec e^{\;\pm}_b)%|_{{}_{\suup^+}}
$.
% \hspace{1cm}
% \bar{g}^-_{ab}\equiv e^{-\mu}_a e^{-\nu}_b g_{\mu\nu}|_{{}_{\suup^-}}=
% (\vec e^{\;-}_a\cdot \vec e^{\;-}_b)%|_{{}_{\suup^-}}
% .
% \]
These
are the first fundamental forms
of $\sup$ inherited from $(\mm2^\pm,g^\pm)$.
Now, as shown in \cite{Clarke-Dray,Mars-Seno}, the necessary and sufficient condition
for the existence of a {\em continuous\/} extension $g$ of the
metric to the whole manifold $\mmm$ 
such that $g|_{{\cal V}^+}=g^+$ and $g|_{{\cal V}^-}=g^-$
is
\be
\bar{g}^+=\bar{g}^-.
\label{prelim}
\ee
These relations,
which can also be expressed as
$ds^{2+}|_{\sigma^+}\stackrel{\sigma}{=}ds^{2-}|_{\sigma^-}$
(where $\eqq$ implies that both sides of the equality must be evaluated on
$\sup$),
are the {\em preliminary junction
conditions} \cite{Mars-Seno}.
Now, the bases $\{\vec e^{\;+}_a\}$ and $\{\vec e^{\;-}_a\}$
can be identified,
\be
% \d \Phi^+\left(\left.\frac{\partial}{\partial \xi^a}\right|_{\suup}\right)=
% \d \Phi^-\left(\left.\frac{\partial}{\partial
\d \Phi^+\left(\frac{\partial}{\partial \xi^a}\right)=
\d \Phi^-\left(\frac{\partial}{\partial
\xi^a}\right),
\label{eq:igualvectan}
\ee
as can the hypersurfaces $\sup^+\equiv \sup^-$, so
henceforth we represent both  $\sup^\pm$ by $\sup$.
Essentially, we are identifying the abstract
manifold $\sup$ with its images $\sup^+=\sup^-$ in $(\mmm,g)$.

In order to impose the remaining junction conditions we need a
one-form, $\bm n$, normal to the hypersurface,
defined through the condition
$
\bm n^\pm(\vec e^{\;\pm}_a)=0.
%\label{eq:normals}
$
Since in the
final matched manifold $\mmm$
the normals are to be identified as a single object,
both must have the same norm.
Also
if $\bm n^+$ is to point $\mmm^+$
outwards, then $\bm n^-$ has to point $\mmm^-$ inwards,
and conversely.
In order to deal with general hypersurfaces,
including spacelike and null hypersurfaces,
we will also need the
rigging vectors $\vec\ell^{\;\pm}$ on $\sup^\pm$ \cite{schouten},
which are defined
as vector fields 
on $\sup^{\pm}$ and transversal 
to
$\sup^{\pm}$\footnote{Note that in the case of non-null hypersurfaces the
normal vector is itself a rigging vector.}.
The riggings are therefore characterised everywhere on $\sup$ by
\be
\bm n^+(\vec \ell^{\;+})\eqq \bm n^-(\vec \ell^{\;-})\neq 0,
\label{eq:normirigg}
\ee
so that the vectors
$\{\vec \ell^{\;\pm},\vec e^{\;\pm}_a\}$ constitute a basis
for the tangent spaces to $\mm2^\pm$ at $\sup^\pm$.
Given that the preliminary conditions allow us to identify
$\{\vec e^{\;+}_a\}$ with $\{\vec e^{\;-}_a\}$,
it only remains to choose the riggings
such that the bases $\{\vec \ell^{\;\pm},\vec e^{\;\pm}_a\}$ 
have the same orientation with
\be
\ell^{+}_{\mu}\ell^{+\mu}\eqq \ell^{-}_{\mu}\ell^{-\mu}, \hspace{1cm}
\ell^{+}_{\mu}e^{+\mu}_{a}\eqq \ell^{-}_{\mu}e^{-\mu}_{a}.
\label{eq:riggings}
\ee
In this way, we can identify the whole 4-dimensional
tangent spaces of $\mm2^\pm$ at $\sup$,
$\{\vec \ell^{\;+},\vec e^{\;+}_a\}\equiv
\{\vec \ell^{\;-},\vec e^{\;-}_a\}\equiv\{\vec \ell,\vec e_a\}$.
We note that if the second equation in (\ref{eq:riggings}) and
the preliminary junction conditions hold then 
the first relation in (\ref{eq:riggings})
is equivalent to (\ref{eq:normirigg})
up to a sign.

The remaining junction conditions amount to the equality
of the generalised second fundamental forms $H^\pm_{ab}$ 
\[
H^{\pm}_{ab} =-\ell^{\pm}_{\nu}e^{\pm\mu}_{a}\nabla^{\pm}_{\mu}
e^{\pm\nu}_{b}.
\]
In the case of non-null hypersurfaces, choosing $\vec{\ell}=
\vec{n}$, the tensors
$H^{\pm}_{ab}$ coincide with the second fundamental forms
$K^{\pm}_{ab} =-n^{\pm}_{\nu}e^{\pm\mu}_{a}
\nabla^{\pm}_{\mu}e^{\pm\nu}_{b}$ inherited by $\sup^{\pm}$ from
$\mm2^\pm$ \cite{Darmois,Clarke-Dray,Mars-Seno}.
Note that the junction conditions
$H^{+}_{ab}\eqq H^{-}_{ab}$
do not depend on the specific choice of the rigging vector \cite{Mars-Seno}.

When symmetries are present, as in
most of the works dealing with spacetime matchings,
one is interested  in the cases
where the matching surface $\sigma$ inherits a particular symmetry of
the two space-times $(\mmm^\pm,g^\pm)$. Such matching is
said to {\em preserve the symmetry}. In practice one demands that the
matching hypersurface is tangent to the orbits of the symmetry group to
be preserved.
A more rigorous definition of matching preserving the
symmetry was recently presented in \cite{Vera-thesis}.
Thus if $(\mmm^+,g^+)$ and $(\mmm^-,g^-)$ both admit a
$m$-dimensional group of symmetries,
the final matched spacetime $(\mmm,g)$ is said to preserve
the symmetry $G_m$ if there exist $m$ vectors on $\sigma$
that are mapped by the push-forwards $d\Phi^+$ and
$d\Phi^-$ to the restrictions of the generators of $G_m$ to
$\sigma^+$ and $\sigma^-$, respectively.
Furthermore, if there is an intrinsically
distinguishable generator of $G_m$ in $\mmm^+$ and $\mmm^-$, 
such as an axial Killing vector,
then the matching preserving the symmetry
must ensure its identification
at $\sup$.

In the cases we shall consider below,
$(\mmm^+,g^+)$ will
correspond to a $G_4$ on $S_3$ LRS spacetime,
thus admitting a cylindrical symmetry (Abelian $G_2$ subgroup),
and $(\mmm^-,g^-)$ to a static OT cylindrically symmetric spacetime.
We shall consider the matching preserving the cylindrical symmetry,
which is represented by an Abelian group $G_2$ \cite{bisch,alancil,Carot-etal}.

%_____________________________________________________________________________
\section{General metric forms with a $G_4$ on $S_3$ which are LRS}
\label{compact-lrs}
%_____________________________________________________________________________
In this section we shall write down in explicit cylindrical-like coordinates
a general compact form of
the metric corresponding to LRS spatially homogeneous spacetimes,
which admit a $G_4$ group of motions on spatial 3-hypersurfaces $S_3$.
We begin by combining the standard metric forms for
all possible $G_4$ on $S_3$ LRS spaces, with $k=\pm 1,0$,
given by \cite{Kramer} (see also \cite{MalcolmLRS})
\bea
\label{11.3}
&ds^2&=-dt^2+a^2(t)dx^2+b^2(t)(dy^2+\Sigma^2(y,k)dw^2)\\
\label{11.4}
&ds^2&=-dt^2+a^2(t)\bm{\sigma}^2_k+b^2(t)(dy^2+\Sigma^2(y,k)dw^2)\\
\label{11.5}
&ds^2&=-dt^2+a^2(t)dx^2+b^2(t)e^{2x}(dy'^2+dw'^2)
\eea
where
\be
\label{sigma}
\Sigma(y,k)=
\left\{
\begin{array}{cl}
\sin y,& k=+1\\
y, & k=0\\
\sinh y, & k=-1
\end{array}
\right.
~~
~~
and
~~
\bm{\sigma}_k  =
\left\{
\begin{array}{cl}
dx+\cos ydw, & k=+1\\
dx-y^2/2 dw, & k=0\\
dx+\cosh ydw, & k=-1
\end{array}
\right.
\ee
Performing a change to polar coordinates $\{y'=y\sin w,
w'=y\cos w\}$
in the line-element (\ref{11.5}),
and following \cite{MalcolmLRS} in a first step,
the above metrics can be combined into a compact form given by
\begin{equation}
\label{comp}
ds^2=-dt^2+a^2(t)\bm{\hat\theta}^2_{nk}+b^2(t)e^{2\epsilon x}(dy^2+
\Sigma^2(y,k)dw^2),
%[(1-\epsilon)\Sigma^2(y,k)+\epsilon]dw^2)
\end{equation}
where
\begin{equation}
\label{theta}
\bm{\hat\theta}_{nk}=dx+nF(y,k)dw,
\end{equation}
\be
\label{F}
F(y,k)=
\left\{
\begin{array}{cl}
-\cos y,& k=+1\\
y^2/2, & k=0\\
\cosh y, & k=-1
\end{array}
\right.
,
\ee
and where $\epsilon$ and $n$ are such that
\begin{equation}
\label{epsi}
\epsilon=0,1;~~n=0,1;~~\epsilon n=\epsilon k=0.
\end{equation}
We note that
\be
\label{Fprime}
\Sigma=F_{,y};~~(\Sigma_{,y})^2+k\Sigma^2=1,
\ee
where here and throughout the comma denotes the 
partial derivative with respect to the indicated variable.
The metrics (\ref{11.3}), (\ref{11.4}) and (\ref{11.5})
are recovered with $\{\epsilon=0,n=0\}$,
$\{\epsilon=0,n=1\}$ and $\{\epsilon=1\}$ respectively\footnote{Note
that for convenience
we have introduced a change in the sign of $w$ in (\ref{11.4})
for the cases $k=0,1$ which results in a change of sign in the expressions
of the Killing vectors as shown in Kramer et al. \cite{Kramer}.}.
The metric (\ref{comp}) with $\epsilon = n=0$ and $k=1$
is the Kantowski-Sachs metric,
which admits no simply transitive $G_3$ subgroup. All
the other cases included in (\ref{comp})
possess a simply transitive $G_3$ subgroup of symmetries
that can be classified according to their Bianchi types:
(\ref{11.4}) corresponds to 
type $II$ for 
$k=0$ and
to types $III, VIII$ and $IX$ for $k\ne 0$; (\ref{11.5}) corresponds to
types $V$ and $VII_h$ and (\ref{11.3})
corresponds to types $I$, $III$ and $VII_0$. These classifications are
summarised in Table
\ref{table:bianchis}.
The cases with $n=0$ admit a multiply transitive $G_3$ on $S_2$.
 
The axial Killing vector associated with the metric (\ref{comp}) is
$\vec\eta_1=\p_w+nk\p_x$, which can be easily shown to define a regular
axis at $y=0$,
that is
%$\vec\eta_1$ obeys
\be
\left.\frac{\nabla_\rho({\vec \eta_1}^2)
\nabla^\rho({\vec \eta_1}^2)}{4{\vec \eta_1}^2}\right.
%\right|_{W_2}\longrightarrow 1.
\label{eq:condregaxis}
\ee
%(\ref{eq:condregaxis}) below for
%$y\to 0$).
% $\nabla_{\alpha}(\eta_1^\beta {\eta_1}_\beta)
% \nabla^{\alpha}(\eta_1^\gamma {\eta_1}_\gamma)/(4\eta_1^\nu {\eta_1}_\nu)
% |_{y\to 0}\to 1$).
tends to 1 as $y\to 0$.
The axial Killing vector $\vec\eta_1$ together with 
$\vec\eta_2=\p_x-\epsilon y \p_y$
generates an Abelian subgroup  $G_2$ on $S_2$.
For the purposes of this paper 
it is desirable to express the metric in 
a form adapted to both $\vec \eta_1$ and $\vec \eta_2$.
%the axial Killing vector.
To do this, we perform the following  coordinate
transformations to cylindrical coordinates
\begin{eqnarray}
\label{transf}
x&\to&z+nk\varphi\nonumber\\
% y&\to&\epsilon r \cos \varphi + r (1-\epsilon)\nonumber\\
% w&\to&\epsilon r \sin \varphi + \varphi (1-\epsilon)\\
y&\to&r e^{-\epsilon z} \nonumber\\
w&\to&\varphi \\
t&\to&t\nonumber
\end{eqnarray}
which
brings the compact metric (\ref{comp}) into the form
\be
\label{compact}
ds^2=-dt^2+a^2(t) \bm{\theta}^2_{nk}
+b^2(t)\left[(dr-\epsilon r dz)^2+
\Sigma(r,k)^2 d\varphi^2\right],
\ee
where
\be
\label{theta1}
\bm{\theta}_{nk}=dz+n(F(r,k)+k)d\varphi.
\ee
To our knowledge, this is a new form of presenting all the
$G_4$ on $S_3$ LRS spacetimes in compact form.
The axial Killing vector is then given by 
\be
\label{akv}
\vec\eta_1=\partial_{\varphi},
\ee
while the other three Killing vectors $\vec\eta_i, ~i=2$ to $4$, are
taken to be
\begin{eqnarray}
\label{killvecs}
\vec\eta_2 &=&\p_z,\nonumber\\
\vec\eta_3 &=&e^{\epsilon z}\left[\sin\varphi \p_r+
\cos\varphi(f(r)\p_\varphi+g(r)\p_z)\right],\\
\vec\eta_4 &=&e^{\epsilon z}\left[\cos{\varphi}\p_r-
\sin\varphi(f(r)\p_\varphi+g(r)\p_z)\right],\nonumber
\label{kkk}
\end{eqnarray}
where we have defined
$f(r)=\Sigma_{,r}/\Sigma$ and $g(r)=n(\Sigma-f(F+k))$.
%____________________________TABLE___________________________________
\begin{table}[!tb]
\begin{center}
\begin{tabular} {|c|c| c| c|}
\hline
Bianchi Types & $\epsilon$&$n$&$k$ \\
\hline
I,III&0&0&-1 
\\
VII$_{0}$&0&0&0
\\
%0&0&1 & not possible
%\\
VIII, IX & 0&1&-1
\\
II & 0&1&0
\\
III & 0&1&1
\\
V,VII$_{h}$&1&0&0
\\
\hline
\end{tabular}
\caption{\label{table:bianchis} Bianchi types of the possible
subgroups $G_3$ on $S_3$ according to the values
$\{\epsilon,k,n\}$ of metric (\ref{comp}).}
\end{center}
\end{table}
%_____________________________________________________________________________
\section{The matching hypersurface}
% in cylindrical symmetry}
\label{parame}
%_____________________________________________________________________________
Our aim is to match a spacetime corresponding to
%exterior given by 
the metric (\ref{compact}) and
a static OT cylindrically symmetric spacetime. %interior.
The metric (\ref{compact}) can be cast in the more general form, say $g^+$, 
given by
\be
\label{non-static}
ds^{2+}=-\hat{A}^2dt^2+\hat{B}^2dr^2-2\epsilon r \BB^2 dr dz
+\hat{C}^2 d{\varphi}^2
+2\hat{E}dzd\varphi+\hat{D}^2d{z}^2,
\ee 
where $\hat{A},\hat{B},\hat{C},\hat{D}$ and $\hat{E}$ are functions
of $t$ and $r$. The line-element (\ref{compact}) is recovered
by making the identifications
\bea
\label{idents}
\hat{A}^2(t,r)&=&1,\nonumber\\
\hat{B}^2(t,r)&=&b^2(t),\nonumber\\
\hat{C}^2(t,r)&=&b^2(t)\Sigma^2(r,k)+n a^2(t)(F(r,k)+k)^2,\\
\hat{D}^2(t,r)&=&a^2(t)+\epsilon r^2 b^2(t),\nonumber\\
\hat{E}(t,r)&=&n a^2(t) (F(r,k)+k).\nonumber
\eea
%and the four corresponding Killing vectors are given by (\ref{akv}) and
%(\ref{kkk}).
The usefulness of writing (\ref{compact})
as (\ref{non-static}) using (\ref{idents})
will become clear in the next section.
In the following we shall take the functions in (\ref{non-static})
to be arbitrary functions, with
%except for the fact that
$$
\epsilon\BE=0,
$$
which follows from (\ref{epsi}).
%, for simplicity, will be used all throughout the next sections.

The metric $g^{-}$
is assumed to be static and cylindrically symmetric \cite{Carot-etal},
admitting, in principle, a maximal group \footnote
{The $G_3$ group is then taken to be Abelian.
Other algebraic types require the existence of
more symmetries \cite{Carot-etal}, and therefore 
need to be studied separately.}
$G_3$ on $T_3$ containing
an Abelian subgroup $G_2$ on $S_2$ which includes an axial
symmetry \cite{bisch,alancil}.
The orbits of this $G_2$ subgroup are also assumed to
generate orthogonal surfaces, i.e. the
group $G_2$ is assumed to act orthogonally transitively (OT).
This is the analogue of the `circularity condition',
usually used within the context of
stationary axisymmetric interior problems,
where it implies non-convectivity
in fluids without energy flux \cite{carter69}.
This assumption is in fact a consequence of the existence
of an axis of symmetry in spacetimes with
certain types of matter content, including vacuum \cite{carter69,Carot-etal}.

Now one can always find a coordinate system
$\{T,\rho,\tilde{\varphi}',\tilde{z}'\}$ adapted to the Killing
vectors $\{\partial/\partial T,
\partial/\partial\tilde{\varphi}', \partial/\partial\tilde{z}'\}$,
where $\p/\p \tpp'$ is the axial Killing vector,
such that the metric $g^-$ is given by
\be 
\label{static}
{d s^{2}}^{-}=- \CA^{2} d T'^{2} + \CB^2 d \rho^{2} 
+ \CC^{2} d \tpp'^{2} + \CD^{2} d \tz'^{2}+ 2 \CE d \tpp' d \tz' ,
\ee
where $\CA, \CB, \CC, \CD$ and $\CE$ are %$c^1$
functions of $\rho$.

Following the matching procedure specified in Section \ref{geom} 
we proceed by specifying the two embeddings $\sigma^{\pm}$. 
The embedding of $\sigma^+$ can be defined 
by choosing the appropriate
coordinates on $\sigma$ denoted by 
\be
\label{coord}
\{ \xi^a\}=\{\lambda,\phi,\zeta \}.
\ee
%(throughout this section
%we follow the notation of Section \ref{geom}).
The coordinate $\phi$ is chosen such that 
the vector field $\p/\p\phi$ is mapped, by
the push-forward $d\Phi^+$, at every point in $\sigma$,  
into the restriction of the
axial Killing vector $\vec\eta_1=\p/\p\varphi$ on $\sigma^+$, that is,
\be
\label{phi1}
d\Phi^+\left(\frac{\p}{\p\phi} \right)=
\left.\frac{\p}{\p\varphi}\right|_{\sigma^+}\equiv \vec{e}^+_2,
\mbox { and thus }
%\frac{\p\Phi^{2+}}{\p\phi}\left.\frac{\p}{\p\varphi}\right|_{\sigma^+}=
%\Longleftrightarrow
\frac{\partial {\Phi^0}^+}{\partial \phi}=
\frac{\partial {\Phi^1}^+}{\partial \phi}=
\frac{\partial {\Phi^3}^+}{\partial \phi}=0,\;
\frac{\partial {\Phi^2}^+}{\partial \phi}=1.
\ee
Since we want the matching to preserve
the cylindrical symmetry (i.e. a $G_2$ on $S_2$
containing the axial symmetry generated by $\vec\eta_1$),
there must exist a vector field $\vec\gamma$ in $\sigma$ which
is mapped to the restriction on $\sigma^+$ of a Killing vector 
that, together with $\vec\eta_1$, generates a $G_2$ on $S_2$.
The only possibility is for this Killing vector to be a linear combination of 
$\vec\eta_1$ and $\vec\eta_2$.
In other words, we have
$d\Phi^+\left( \vec\gamma\right) =
\left. a\vec\eta_1\right|_{\sigma^+}+
\left. b\vec\eta_2\right|_{\sigma^+}$ where $a$ and $b$
are arbitrary constants with $b\neq 0$.
We can now use the fact that the group which is preserved
is automatically inherited by the hypersurface $\sigma$ \cite{Vera-thesis,Vera01}
in which it has the same algebraic type \cite{Vera01}.
Since the $G_2$ generated by $\vec\eta_1$ and $\vec\eta_2$
is Abelian then the vectors $\p/\p\phi$ and $\vec\gamma$,
which are Killing vectors in $\sigma$, commute. 

We can now choose
a coordinate $\zeta$ such that $\vec\gamma=\p/\p \zeta$ and use a coordinate
transformation
$\zeta'=b\,\zeta$, $\phi'=\phi+
a\,\zeta$ in $\sup$ in order to obtain (after dropping the primes)
$$
\d \Phi^+\left(\frac{\p}{\p \zeta}\right)
=\left. \frac{\p}{\p z}\right|_{\suup^+} \equiv \vec e^{\;+}_3,
%\Longleftrightarrow
$$
and thus
\be
\frac{\partial {\Phi^0}^+}{\partial \zeta}=
\frac{\partial {\Phi^2}^+}{\partial \zeta}=
\frac{\partial {\Phi^1}^+}{\partial \zeta}=0,~
\frac{\partial {\Phi^3}^+}{\partial \zeta}=1,
\label{phi2}
\ee
leaving (\ref{phi1}) unchanged.
Finally, the remaining coordinate $\lambda$ can always
be chosen such that the image of
$\p/\p\lambda$ through $\d \Phi^+$
is orthogonal to $\vec{e}^+_2$ and $\vec{e}^+_3$.
This implies
$\partial {\Phi^2}^+/\partial \lambda\eqq
-\BE/\BC^2\partial {\Phi^3}^+/\partial \lambda$ and
$$
% &&\frac{\partial {\Phi^2}^+}{\partial \lambda}\eqq
% -\frac{\BE}{\BC^2}
% \frac{\partial {\Phi^3}^+}{\partial \lambda},\\
%&&
\frac{\partial {\Phi^3}^+}{\partial \lambda}\eqq \epsilon {\Phi^1}^+
\frac{\BB^2\BC^2}{\deltt}
%(\BC^2\BD^2-\BE^2)^{-1}
\frac{\partial {\Phi^1}^+}{\partial \lambda},
$$
where 
$\deltt\equiv \BC^2\BD^2-\BE^2$,
and thus
%Making implicit use of (\ref{idents})
%and the fact that $\epsilon n=0$ 
%and so, for instance, $n e^{-2\epsilon z}=n$,
$$
\frac{\partial {\Phi^2}^+}{\partial \lambda}=0.
% \frac{\partial {\Phi^3}^+}{\partial \lambda}\eqq
% \epsilon {\Phi^1}^+ e^{2\epsilon {\Phi^3}^+}\left(\frac{b}{a}\right)^2
% \frac{\partial {\Phi^1}^+}{\partial \lambda}.
%\label{coicoi}
$$
Hence we obtain the following expression for $\vec{e}^+_1$
\be
\label{phi3}
d\Phi^+\left(\frac{\p}{\p\lambda} \right)\eqq
\frac{\p\Phi^{0+}}{\p\lambda}\left.\frac{\p}{\p t}\right|_{\sigma^+}+
\frac{\p\Phi^{1+}}{\p\lambda}\left.\frac{\p}{\p r}\right|_{\sigma^+}
+\epsilon {\Phi^1}^+ \frac{\BB^2\BC^2}{\deltt}
\frac{\partial {\Phi^1}^+}{\partial \lambda}
\left.\frac{\p}{\p z}\right|_{\sigma^+}
\equiv \vec{e}^+_1.
\ee
By denoting the embedding
$\{\Phi^{0+},\Phi^{1+},\Phi^{2+},\Phi^{3+}\}$
in (\ref{embed}) as $\{t,r,\varphi,z\}$, 
the matching surface $\sigma^+$ is parametrized as
\be
\label{surf+}
\sigma^+=\{(t,r,\varphi,z): t=t(\lambda),
r=r(\lambda),\varphi=\phi,z=\zeta+f_{z}(\lambda) \},
\ee
where $t(\lambda)$ and $r(\lambda)$
are functions of $\lambda$ restricted
by the fact that $d\Phi^+$ has to be of rank 3, that is
$\dot t^2+\dot r^2\neq 0$,
and
\be
\dot f_{z}(\lambda)\eqq\epsilon r(\lambda)\dot r(\lambda)\frac{\BB^2\BC^2}
{\deltt},
\label{dotfz}
\ee
where the dot denotes differentiation with respect to $\lambda$.
For $\epsilon=0$ we can choose $f_{z}=0$ without loss
of generality.

We now consider the embedding $\Phi^-$. The preservation
of the cylindrical symmetry implies that the
axial Killing vectors
from both sides
have to coincide at the matching hypersurface.
Since the tangent spaces of both $\sigma^+$ and $\sigma^-$
are going to be identified (see Section (\ref{geom})),
we only have to impose
%which gives
% \be
% \label{dphi-}
% d\Phi^-\left(\frac{\p}{\p\phi}\right)=
% \left.\frac{\p}{\p\tilde\varphi'}\right|_{\sigma^-}\equiv \vec{e}^-_2.
% \ee
$
d\Phi^- (\p/ \p\phi)=
\p/\p \tilde\varphi'|_{\sigma^-}\equiv \vec{e}^-_2.
$
The image of the vector
$\partial/\partial \zeta$ must complete the basis
of some $G_2$ on $S_2$ subgroup of
the $G_3$ on $T_3$ admitted by $(\mm2^-,g^-)$.
Therefore the image of $\p/\p\zeta$ takes the form 
\be
\label{general-form}
d\Phi^-\left(\frac{\p}{\p\zeta}\right)=
a \left.\frac{\p}{\p \tpp'} \right|_{\suup^-}+
b \left.\frac{\p}{\p \tilde z'}\right|_{\sigma^-}+
c \left.\frac{\p}{\p T'}\right|_{\sigma^-}\equiv \vec e^{\;+}_3,
\ee
where $a, b$ and $c$ are arbitrary constants, such that the inequality
\be
\left.\left[-c^2\CA^2\CC^2+b^2(\CC^2\CD^2-\CE^2)\right]\right|_{\suup^-}>0
\label{eq:onS2}
\ee
gives the necessary and sufficient condition
for having spacelike orbits, implying $b\neq 0$.
Now condition (\ref{eq:onS2}) will be automatically
satisfied
%(if $b\neq 0$)
once the preliminary junction conditions are fulfilled,
since $\vec e^{\;+}_3$ is always spacelike, which makes
$\vec e^{\;-}_3$ spacelike. 

In order to simplify (\ref{general-form}), we perform a change in $(\mm2^-,g^-)$ to a
new
coordinate system $\{T,\rho,\tpp,\tz\}$ defined by
$$
T=T'-\frac{c}{b}\tz',~
\tpp=\tpp'-\frac{a}{b}\tz',~
\tz=\frac{1}{b}\tz',
$$
which is still adapted to the axial Killing vector, that is,
$\partial/\partial \tpp=\partial/\partial \tpp'$.
In these new coordinates the line-element for $\g^-$ reads
\be
\label{hola2}
ds^{2-}=
-A^2dT^2-2cA^2dTd\tilde z+B^2d\rho^2+C^2 d\tilde{\varphi}^2+2Ed\tilde\varphi
d\tilde z+
D^2d\tilde{z}^2,
\ee
where all the `non--hatted' functions depend only on $\rho$ and
we have set $A\equiv \CA$, $B\equiv \CB$, $C\equiv \CC$ and
\be
D^2\equiv -c^2 \CA^2+a^2 \CC^2+b^2\CD^2+2ab\CE,\hspace{1cm}
E\equiv a\CC^2+b\CE.
\label{eq:canvfuncions}
\ee
Note that this change of coordinates is well
defined whenever (\ref{eq:onS2}) holds, since this implies
$C^2D^2-E^2>0$.
Now, taking the embedding $\Phi^{\alpha -}$ in the
new coordinates $\{T,\rho,\tpp,\tz\}$,
the images of $\partial/\partial \phi$ and $\partial/\partial \zeta$
simplify to
\be
\d \Phi^-\left(\frac{\p}{\p\phi} \right)
=\left.\Partp \right|_{\suup^-}\equiv\vec e^{\;-}_2,
\mbox { with }
\frac{\partial {\Phi^0}^-}{\partial \phi}=
\frac{\partial {\Phi^1}^-}{\partial \phi}=
\frac{\partial {\Phi^3}^-}{\partial \phi}=0,\;
\frac{\partial {\Phi^2}^-}{\partial \phi}=1
\label{eq:phi-}
\ee
\be
\d \Phi^-\left(\Parsz\right)
=\left.\Partz \right|_{\suup^-}\equiv \vec e^{\;-}_3,
\mbox { with }
\frac{\partial {\Phi^0}^-}{\partial \zeta}=
\frac{\partial {\Phi^1}^-}{\partial \zeta}=
\frac{\partial {\Phi^2}^-}{\partial \zeta}=0,\;
\frac{\partial {\Phi^3}^-}{\partial \zeta}=1.
\label{dphi-2}
\ee
Since the image of $\p/\p \lambda$ by $d\Phi^+$ is orthogonal
to $\vec{e}_2^+$ and $\vec{e}_3^+$,
then its image by $d\Phi^-$, i.e. $\vec{e}_1^-$,
must be orthogonal to $\vec{e}_2^+$ and $\vec{e}_3^+$,
essentially because of the preliminary junction conditions, thus resulting in 
\be
\frac{\partial {\Phi^3}^-}{\partial \lambda}
\eqq\frac{cA^2C^2}{\delt}\frac{\partial {\Phi^0}^-}{\partial \lambda},
\hspace{1cm}
\frac{\partial {\Phi^2}^-}{\partial \lambda}\eqq
-\frac{E}{C^2}\frac{\partial {\Phi^3}^-}{\partial \lambda}
\label{eq:embed-b},
\ee
where we have defined $\delt\equiv C^2D^2-E^2$,
so that we have
\bean
&&\d \Phi^-\left(\Parsl \right)=\hspace{7cm}\\
&&\left.\frac{\partial {\Phi^0}^-}{\partial \lambda} \Partt\right|_{\suup^-}+
\left.\frac{\partial {\Phi^1}^-}{\partial \lambda}\Parrho\right|_{\suup^-}-
\frac{c A^2}{\delt}
\left.\frac{\partial {\Phi^0}^-}{\partial \lambda}
\left(E\Partp-C^2 \Partz \right) \right|_{\suup^-}\equiv \vec e^{\;-}_1.
\eean
As a result the matching surface $\sigma^-$ is parametrised by 
\be
\label{surf-}
\sigma^-=\{(T,\rho,\tilde\varphi,\tilde z):
T=T(\lambda), \rho=\rho(\lambda),\tilde\varphi=\phi+
f_{\tilde\varphi}(\lambda),\tilde z=\zeta+f_{\tilde z}(\lambda) \},
\ee
where $T(\lambda)$ and $\rho(\lambda)$ are arbitrary
functions of $\lambda$,
and $f_{\tilde\varphi}(\lambda)$ and $f_{\tilde z}(\lambda)$
are related to $T(\lambda)$ by (\ref{eq:embed-b}).
In this way, we have obtained the most general
parametrisations for 
the interior and exterior matching surfaces $\sigma^+$ and $\sigma^-$.

 The matching procedure described above involves some free parameters
 in the metric $\g^-$ which account for the possible inequivalent
 ways the two spacetimes $(\mm2^\pm,\g^\pm)$
 can be joined \cite{MASEuni,Vera-thesis}.
 %with metrics (\ref{eq:ds2nscil}) and (\ref{eq:ds2scil}).
 Some of these parameters may encode the freedom in choice of coordinates,
 while others may imply physical differences \cite{MASEuni}.

 Thus, in practical situations where the metric
 (\ref{static}) is given, we must use
 the new `non-hatted' functions obtained from
 the relations (\ref{eq:canvfuncions}), which then depend on
 the parameters $a, b$ and $c$, in the equations arising
 from the junction conditions.
 On the other hand, if the spacetime metric $\g^-$ is unknown,
then according to the way we have set up the problem, 
the matching problem must be treated using (\ref{hola2}).
 In that case, to recover the form (\ref{static}),
 we need to invert equations (\ref{eq:canvfuncions})
 in order to obtain the hatted functions appearing 
 in (\ref{hola2}).
 Only after this inversion is performed is it possible to determine
 whether different values of the parameter $c$
 correspond to different ways of joining the spacetimes
 give rise to equivalent matchings.

 %-------------------------------------------------------
 \section{Junction conditions}
 \label{junction}
 %-------------------------------------------------------
 We recall that in order to derive 
 the junction conditions we have to calculate the first and second
 fundamental forms
 for both $\sigma ^+$ and $\sigma^-$.
 For the $g^+$ metric (\ref{non-static}),
 the parametric form of $\sigma^+$ (\ref{surf+}) gives
 $dt|_{\sigma^+}=\dot{t}d\lambda,~ dr|_{\sigma^+}=\dot{r}d\lambda,~
 d\varphi|_{\sigma^+}=d\phi$ and 
 $dz|_{\sigma^+}=d\zeta+\dot f_{z}d\lambda.$
 Using (\ref{dotfz}), the first fundamental form
 on $\sigma^+$
 can be written as
 \be
 \label{hola1}
 ds^{2+}|_{\sigma^+}\eqq (-\hat{A}^2\dot{t}^2+\bbb^2\dot{r}^2)d\lambda^2
 +\hat{C}^2 d\phi^2+2\hat{E}d\phi d\zeta 
 +\hat{D}^2 d\zeta^2,
 \ee
 where 
 $$
 \bbb^2\equiv \BB^2\left(1-\epsilon r^2
 \frac{\BB^2\BC^2}{\deltt}\right),
 $$
 which can be reduced to $\BB^2(1-\epsilon r^2
 \BB^2/\BD^2)$ making use of $\epsilon \BE=0$.
 %(\ref{idents}) and the fact that $n\epsilon=0$.
 Similarly for the $g^-$ metric (\ref{hola2}),
 the first fundamental form %$\bar g^-_{ab}$
 on $\sigma^-$ (\ref{surf-}) is given by
 \be
 ds^{2-}|_{\sigma^-}\eqq (-{\cal A}^2\dot T^2+B^2 \dot \rho^2)d\lambda^2
 +C^2d\phi^2+2E d\zeta d\phi+D^2 d\zeta^2,
 \label{hola7}
 \ee
 where 
 $$
 {\cal A}^2=A^2\left(1+c^2\frac{A^2C^2}{\delt}\right).
 $$
 The equality of the first fundamental forms 
 (\ref{hola1}) and (\ref{hola7}) gives 
 \bea
 \label{pre1}
 -\BA^2 \dot{t}^2+\bbb^2\dot{r}^2  & \stackrel{\sigma}{=}& -{\cal
 A}^2\dot{T}^2+B^2\dot{\rho}^2\\
 \label{pre2}
 \hat{D}& \stackrel{\sigma}{=}& D\\
 \label{pre3}
 \hat{C}&\stackrel{\sigma}{=}& C\\
 \label{pre4}
 \hat{E} &\stackrel{\sigma}{=}& E.
 \eea
 In order to derive the remaining
 junction conditions
 we need the normal forms to $\sigma^\pm$,
 which can be written as
 \be
 \label{hola14}
 \bm{n}^+=\hat{A}\bbb(-\dot{r}dt+\dot{t}dr)|_{\sigma^+},
 ~~~~\bm{n}^-=\gamma{\cal
 A}B
 (-\dot\rho dT+\dot{T}d\rho)|_{\sigma^-}
 \ee
 so that they have the same norm on $\sigma$
 %(using (\ref{hola10}))
 and where $\gamma=\pm 1$ defines the two possible relative
 orientations.
 The rigging vectors can be obtained from 
 (\ref{eq:normirigg}) and (\ref{eq:riggings}), and a suitable
 choice is
 \bea
 \label{hola15}
 &&\left.\vec{\ell}^{+}= -\frac{\dot{r}}{\BA^{2}}\frac{\p}{\p t}
 \right|_{\sigma^{+}}+
 \left.\frac{\dot{t}}{\bbb^{2}}\frac{\p}{\p r} \right|_{\sigma^{+}},\\
 % \hspace{1cm}
 \label{hola16}
 &&\left.\vec{\ell}^{-}=G \left[-\alpha^{2} \frac{\dot{\rho}}{{\cal
 A}^{2}}\frac{\p}{\p t}
 +\frac{\dot{T}}{B^2}\frac{\p}{\p\rho}
 +\frac{1}{\delt}\left(\alpha^2 \dot{\rho}\frac{cA^2}{{\cal A}^2}
 +\dot t\frac{\epsilon r\BB^2}{\bbb^2 G}\right)
 \left(E\frac{\p}{\p \tpp}-C^2\frac{\p}{\p \tz} \right)
 \right] \right|_{\sigma^{-}}
 \eea
 %where we have defined $\delt\equiv\BC^2\BD^2-\BE^2,(\eqq C^2 D^2-E^2)$
 %because of (\ref{pre2})-(\ref{pre4}), and
 where $G\neq 0$ and $\alpha$ are functions that satisfy
 \bea
 \label{Galpha1}
 &&\frac{1}{\BA\bbb}\left(\bbb^2\dot r^2+\BA^2\dot t^2\right)\eqq
 \frac{\gamma G }
 {\aaa B}\left(\alpha^2B^2\dot \rho^2+\aaa^2\dot T^2\right),
 \\
 \label{Galpha2}
 &&2 \,\dot{r} \,\dot{t} \eqq G \left(\alpha^{2} + 1\right)\,
 \dot{T} \dot{\rho}.
 \eea
 %in order to comply with (\ref{eq:normirigg}) and (\ref{eq:riggings}).
 The explicit expressions for the
 junction conditions $H^+_{ab}\stackrel{\sigma}{=}H^-_{ab}$ can be
 written, using (\ref{pre2})-(\ref{pre4}), in the forms
 %H^{}_{\phi\phi}$ and $H^{}_{\zeta\zeta}$ and their explicit form reads
 \bea
 H_{\lambda\lambda}:\hspace{.1cm}
 \dot{r}\,\ddot{t}+\dot{t}\,\ddot{r}+\dot{r}\,\left(\frac{\BA_{,t}}{\BA}
 \dot{t}^{2}+2 \frac{\BA_{,r}}{\BA}\dot{r}\,\dot{t}+
 \frac{\bbb\bbb_{,t}}{\BA^{2}}
 \dot{r}^{2}\right)+ \dot{t}\,\left(\frac{\bbb_{,r}}{\bbb}
 \dot{r}^{2}+2 \frac{\bbb_{,t}}{\bbb}\dot{r}\,\dot{t}+
 \frac{\BA\BA_{,r}}{\bbb^{2}}
 \dot{t}^{2}\right)  \nonumber \\
 \eqq G\left\{\alpha^{2}\dot{\rho}\,\ddot{T}+\dot{T}\ddot{\rho}+2\,
 \alpha^{2}
 \dot{\rho}^{2}\dot{T}\frac{\aaa_{,\rho}}{\aaa}+\dot{T}\left[\dot{\rho}^{2}
 \frac{B_{,\rho}}{B}+\dot{T}^{2}\frac{\aaa\aaa_{,\rho}}{B^{2}}
 \right]\right\},
 \hspace{.5cm}\label{eq:h11}
 \eea
 \bea
 &&H_{\lambda\phi}:\hspace{.5cm} 0\eqq
 c\,\left(2EC_{,\rho}-E_{,\rho}C\right) \label{eq:h12}\\
 &&H_{\lambda\zeta}:\hspace{.5cm}
 -\frac{\gamma \epsilon}{\BA \bbb}\left(\frac{\bbb^2}{\BB^2}\right)_{,t}
 \eqq \frac{r}{\aaa B c}\left(\frac{\aaa^2}{A^2}\right)_{,\rho}
 \label{eq:h13}
 \eea
 \bea
 &&H_{\phi\phi}:\hspace{.5cm}
 \dot{r}\frac{\BC_{,t}}{\BA^{2}}-\dot{t}\frac{\BC_{,r}}{\bbb^{2}} \eqq
 -G\dot{T}\frac{C_{,\rho}}{B^{2}}, \label{eq:h22}\\
 &&H_{\zeta\zeta}:\hspace{.5cm}
 \dot{r}\frac{\BD_{,t}}{\BA^{2}}-\dot{t}\frac{\BD_{,r}}{\bbb^{2}} \eqq
 -G\dot{T}\frac{D_{,\rho}}{B^{2}}, \label{eq:h33}\\
 &&H_{\phi\zeta}:\hspace{.5cm}
 \dot{r}\frac{\BE_{,t}}{\BA^{2}}-\dot{t}\frac{\BE_{,r}}{\bbb^{2}} \eqq
 -G\dot{T}\frac{E_{,\rho}}{B^{2}}, \label{eq:h23},
 \eea
 where in (\ref{eq:h12})-(\ref{eq:h13})
 we have used the fact that, since
 $\epsilon \BE=0$, then $\epsilon E_{,\rho}\dot{\rho}\eqq0$.
 Note that although there is a factor $c^{-1}$ in (\ref{eq:h13}),
 the right hand side of this equation vanishes identically
 for $c=0$.
The usefulness of using (\ref{non-static}) for (\ref{compact})
becomes clear from the symmetry of the above equations.

 With the exception of equation (\ref{eq:h13}), the set of junction conditions 
 (\ref{pre1})-(\ref{pre4}) and (\ref{Galpha1})%, (\ref{Galpha2})
 %and (\ref{eq:h11})
 -(\ref{eq:h33}),
 is formally the same as those given in
 \cite{Seno-Vera,Vera-thesis},
 where the conditions for a
 matching preserving the $G_2$ symmetry of two
 OT cylindrically symmetric spacetimes, one assumed to be static, were studied.
 Following \cite{Vera-thesis}, the combination of (\ref{pre2})-(\ref{pre4}),
 their derivatives along $\sigma$,
 (\ref{Galpha2}) and
 (\ref{eq:h22})-(\ref{eq:h23}) lead to 
 \begin{itemize}
 \item{(i)} $\BC_{,t}\eqq\BC_{,r}\eqq 0 \Longleftrightarrow C_{,\rho} \eqq 0$.
 \item{(ii)} $\BD_{,t}\eqq\BD_{,r}\eqq 0 \Longleftrightarrow D_{,\rho} \eqq 0$.
 \item{(iii)} $\BE_{,t}\eqq\BE_{,r}\eqq 0 \Longleftrightarrow E_{,\rho} \eqq 0$.
 \item{(iv)} $\dot{r}=0 \Longleftrightarrow \dot{\rho}=0, \hspace{3mm}$
 and then necessarily $\BC_{,t}= \BD_{,t} = \BE_{,t} = 0$.
 \item{(v)} $\dot{t}=0 \Longleftrightarrow \dot{T}=0, \hspace{3mm}$
 and then necessarily $\BC_{,t}\eqq \BD_{,t} \eqq \BE_{,t} \eqq 0$.
 \end{itemize}
 It must be stressed that due to (iv) we cannot have a matching
 across $\sigma$ defined by $\dot\rho=0$, and equivalently
 by $\dot r=0$,
 since from
 (\ref{idents}) this would imply a static
 LRS region. We summarize this latter result in the following Lemma:
 \begin{lemma}
 \label{singler}
 A non-static $G_4$ on $S_3$ LRS spacetime (\ref{compact})
 cannot be matched
 to a static OT cylindrically symmetric spacetime (\ref{static})
 across a hypersurface with
 $\dot \rho=0$ (or equivalently $\dot r=0$),
 preserving the cylindrical symmetry.\fin
 \end{lemma}
Furthermore, the combination of equations that led to the previous
statements, also
imply the following important conditions on $\sigma$:
 \be
 \label{exterior}
 (I) ~\hat{D}_{,t}\hat{C}_{,r}-\hat{D}_{,r}\hat{C}_{,t}\stackrel{\sigma}{=}0,~~~
 (II)~ \hat{E}_{,t}\hat{D}_{,r}-\hat{E}_{,r}\hat{D}_{,t}
 \stackrel{\sigma}{=}0,~~~
 (III)~ \hat{E}_{,t}\hat{C}_{,r}-\hat{E}_{,r}\hat{C}_{,t}\stackrel{\sigma}{=}0.
 \ee
 These are the so-called {\em exterior\/} conditions,
 which led to the impossibility of the
 cylindrically symmetric analogues to the Einstein-Straus model
 in \cite{Seno-Vera,Vera-thesis}. We emphasise that these conditions involve
 only the coefficients of the $g^+$ metric. Three possibilities may arise:
 (a) they cannot be satisfied, and thus
the matching is impossible,
(b) they impose constraints 
on the matching, and in fact determine $\sigma^+$ if the functions
$\BC,\BD,\BE$ are given,
and (c) they are satisfied automatically and 
therefore give no information. 

Finally, following \cite{Seno-Vera,Vera-thesis} we find that,
 after the substitution of $G$ and $\alpha$, the complete set of matching
 conditions can be written as
 \be
 \label{hola19}
 C_{,\rho}\dot{T}\frac{\aaa}{B} \eqq \gamma
 \left(\BA\frac{\BC_{,r}}{\bbb} \dot{t}
 +\bbb\frac{\BC_{,t}}{\BA} \dot{r}\right), \label{eq:new}
 \ee
 \be
 \label{hola20}
 C_{,\rho}^{2} B^{-2}\eqq
 \frac{\BC_{,r}^{2}}{\bbb^2}-
 \frac{\BC_{,t}^{2}}{\BA^2}, \label{eq:crho}
 \ee
 \bea
 \label{hola21}
 \dot{T} C_{,\rho}^{2} \aaa_{,\rho} B^{-3}
 %\left(1+\frac{c^2}{c_2^2}\right)^{1/2}
 \eqq 
 \left(\frac{\BC_{,r}^{2}}{\bbb^2}-\frac{\BC_{,t}^{2}}{\BA^2}\right)
 \left(\frac{\BA_{,r}}{\bbb}\dot{t}+\frac{\bbb_{,t}}{\BA}\dot{r}\right)
 -\frac{\BC_{,t}}{\BA}\frac{\BC_{,r}}{\bbb}\left(\frac{\bbb_{,t}}{\bbb}\dot{t}
 +\frac{\bbb_{,r}}{\bbb}\dot{r}\right) \nonumber \\
 +\frac{\BC_{,t}}{\BA}\frac{\BC_{,r}}{\bbb}
 \left(\frac{\BA_{,t}}{\BA}\dot{t}+\frac{\BA_{,r}}{\BA}\dot{r}\right)-
 \frac{\BC_{,r}}{\bbb}\left(\frac{\BC_{,tt}}{\BA}\dot{t}+
 \frac{\BC_{,tr}}{\BA}\dot{r}\right)+
 \frac{\BC_{,t}}{\BA}\left(\frac{\BC_{,tr}}{\bbb}\dot{t}+
 \frac{\BC_{,rr}}{\bbb}\dot{r} \right), \label{eq:arho}
 \eea
plus the analogous forms for $D$ and $E$
(i.e. changing $C$ by
$D$ and $E$ respectively in the expressions (\ref{eq:new})-(\ref{eq:arho})),
together with (\ref{pre2})-(\ref{pre4}),
 (\ref{eq:h12}), (\ref{eq:h13}) and the exterior
 conditions (\ref{exterior}). Note that not all these equations are
 independent since the set of equations for $C(\rho)$,
 (\ref{eq:new})-(\ref{eq:arho}),
 is related to their analogues for $D$ and $E$ by
 $$
 \BD_{,t} C_{,\rho}\eqq \BC_{,t} D_{,\rho},~~
 \BE_{,t} C_{,\rho}\eqq \BC_{,t} E_{,\rho},~~
 \BE_{,t} D_{,\rho}\eqq \BD_{,t} E_{,\rho}.
 $$

 %_____________________________________________________________________________
 \subsection{The explicit conditions}
 \label{matchLRS}
 %____________________________________________________________________________
 We shall now 
 study the explicit
 matching conditions across non-space-like hypersurfaces
 (so that neither $\dot t$ nor $\dot T$ can
 vanish on $\sigma$) 
 for the $LRS$ homogeneous spacetimes %given by 
 (\ref{compact}) by substituting the metric functions (\ref{idents})
 in the above junction conditions.
 We start by considering the exterior conditions (\ref{exterior}).
 Condition (I) 
 implies that the only possible matchings are those satisfying
 \be
 \label{ext1}
 (a_{,t}\stackrel{\sigma}{=}0)\vee\left(b^2\Sigma_{,r}+a^2n(F+k)
 \stackrel{\sigma}{=}0 \right).
 \ee
 Similarly, condition (II) gives 
 \be
 \label{ext2}
 (a_{,t}\stackrel{\sigma}{=}0)\vee(n=0).
 \ee
 Combining conditions (\ref{ext1}) 
 and (\ref{ext2}) and
 excluding matchings that hold only across a single value for $r$
 corresponding to $\Sigma_{,r}\eqq 0$ (see Lemma \ref{singler}),
 we find that  
 \be
 \label{a}
 a_{,t}\stackrel{\sigma}{=}0
 \ee
 is a {\it necessary condition} for the required matching which, since
 $\dot t\neq 0$ on $\sigma$, implies $a_{,t}=0$ and thus
 \be
 \label{aconst}
 a(t)={\rm constant}(\equiv \beta).
 \ee 
 Considering the remaining
 exterior condition (III) and assuming (\ref{a}), we find 
 \be
 \label{hola30}
 (b_{,t}\stackrel{\sigma}{=}0)\vee(n= 0).
 \ee
 As a result, for $n\ne 0$, a non-static metric (\ref{compact})
 cannot be matched to the static
 % cylindrically static
 metrics (\ref{non-static}) across a non-space-like hypersurface.

 Since we are interested in a non-static LRS region, we shall concentrate
 on the $n=0$ case.
 In this case 
 the matching across a non-space-like $\sigma$ is in principle possible for
 $a=$const and the LRS metric coefficients take the form
 \begin{equation}
 \label{outcome}
 \hat{A}=1,~~\hat{B}=b(t),~~\hat{C}=b(t)\Sigma(r,k),~~\hat{D}^2=\beta^2+
 \epsilon r^2 b^2(t) (=\beta^2+\epsilon \BC^2),~~\hat{E}=0.
 \end{equation}
 Regarding the static region we find that 
 the preliminary junction conditions
 (\ref{pre2})-(\ref{pre4}) for $\dot \rho\neq 0$,
 together with (\ref{outcome}),
 imply that {\em in a neighbourhood of $\sigma$}, the coefficient
 $D(\rho)$ is uniquely determined
 %{\em in a neighbourhood of $\sigma$}
 in terms of $C(\rho)$ by
 \be
 \label{D}
 D^2(\rho)=\beta^2+\epsilon C^2(\rho),
 \ee
 and that
 \be
 \label{E}
 E(\rho)=0.
 \ee
 Furthermore, from statement (i) above we cannot have
 $C_{,\rho}\eqq 0$, since otherwise the only possible LRS regions
 would have to be static. 
 Since we can set $B^2=1$ using the freedom to choose $\rho$,
 then only $A$ and $C$ %(or better $\aaa$ and $C$)
 remain free in (\ref{hola2}).
 Using (\ref{outcome})-(\ref{E}),
 the complete set of junction conditions
 %relating the interior and exterior metrics as well as defining $\sigma$,
 translates into
 \bea
 \label{hola31}
 &&C\eqq b \Sigma,\nonumber\\
 &&C_{,\rho}\dot{T}%\frac{\aaa}{B}
 \aaa
 \stackrel{\sigma}{=}\gamma\frac{\beta}{\sqrt{\beta^2+\epsilon r^2 b^2}}
 \left[\Sigma_{,r}\left(1+\epsilon
 r^2\frac{b^2}{\beta^2}\right)\dot{t}+bb_{,t}\Sigma\dot{r}\right],
 \nonumber\\
 &&%\frac{C^2_{,\rho}}{B^2}
 C^2_{,\rho}
 \stackrel{\sigma}{=}\left(1+\epsilon r^2\frac{b^2}{\beta^2}\right)
 \Sigma_{,r}^2-\Sigma^2 b_{,t}^2,\\
 &&C^2_{,\rho}\dot{T}%\frac{\aaa_{,\rho}}{B^3}
 \aaa_{,\rho}
 \stackrel{\sigma}{=}-\beta\Sigma
 \left\{
 \left[\sqrt{\beta^2+\epsilon r^2 b^2}\Sigma_{,r}b_{,tt}+
 \epsilon r^2 b b_{,t}^2\right] \dot t
 %\right.\nonumber\\
 %&&\hspace{1cm}\left.
 +\frac{\beta^2 b_{,t}\Sigma}
 {\left(\beta^2+\epsilon r^2 b^2\right)^{3/2}}
 (b_{,t}^2+k)^2\dot r
 \right\}\nonumber,
 \eea 
 plus %the relation coming from
 (\ref{eq:h13}), which now explicitly reads
 \be
 \label{h13exp}
 \frac{2\epsilon \gamma \beta r}
 {\left(\beta^2+\epsilon r^2 b^2\right)^{3/2}}b_{,t}
 \eqq
 %\frac{c}{\aaa B}
 \frac{c}{\aaa}\left(\frac{A^2}{\beta^2+\epsilon C^2}\right)_{,\rho}.
 \ee
 This equation shows that when $\epsilon=1$ we also need 
 $c\neq 0$
 in order to have a non-static LRS part. 
 Note that the degree of freedom (i.e. $c$) introduced by the matching
 in (\ref{hola2})
 allows the LRS metrics (\ref{compact}) with $\epsilon=1$ 
 to be matched to static OT
 cylindrically symmetric spacetimes.
 On the other hand if $\epsilon=0$ we either have 
 $c=0$ or $A=$constant. But if $A=$constant, the static region
 (\ref{hola2}) with (\ref{D}) and (\ref{E})
 admits at least one more isometry (see \cite{Vera-thesis}),
 and the matching procedure 
 would then require a different treatment from the outset.

 We also note that given the LRS region, % at the other side,
 i.e. given $b(t)$,
 the static region is not uniquely specified by this matching. This
 is due to the fact that the exterior conditions
 are in this case identically satisfied and hence, as mentioned above,
 they do not prescribe
the matching 
 hypersurface $\sigma^+$ (i.e. $t(\lambda)$ and
 $r(\lambda)$).

 %_______________________________
 \section{Consequences of the matching conditions}
 \label{sec:lemmas}
 %_________________________________
 We briefly summarise the results obtained in subsection \ref{matchLRS}.
 The condition $n=0$ rules out the metric forms (\ref{11.4}) which
 include the Bianchi types
 II, VIII and IX, where the later is an anisotropic generalization of
 FLRW $k=+1$ metrics. From (\ref{hola30}) one therefore has
 \begin{prop}
 \label{llrs2} A non-static $G_4$ on $S_3$ LRS spacetime
 admitting a simply transitive subgroup
 $G_3$ of Bianchi types II, VIII or IX cannot be
 matched to an OT cylindrically
 symmetric static spacetime
 across a non-space-like hypersurface preserving the
 cylindrical symmetry.\fin
 \end{prop}
 Therefore we are left with the metric forms (\ref{11.3}) and
 (\ref{11.5}).
 Now (\ref{11.3}) with $k=+1$
 is the Kantowski--Sachs metric.
 The cases corresponding to $k=0$ and $k=-1$ for $\epsilon=0$
 include Bianchi types III,
 I and VII$_0$, while
 the case $\epsilon=1$ includes Bianchi
 types V and VII$_h$.
These models could be of cosmological interest since
the former generalise $k=0$ FLRW and the latter the $k=-1$ FLRW metrics.
As shown in the previous section, the LRS metric coefficients
 were severely restricted to (\ref{outcome})
 and the
 possible resulting metrics can be summarized as follows:
 %according to the results in the previous section and Lemma (\ref{needaxis})
 \begin{theorem}
 \label{loi}
 The only possible non-static $G_4$ on $S_3$ LRS spacetimes
 that can be matched to an OT cylindrically
 symmetric static spacetime across a non-space-like hypersurface preserving
 the cylindrical symmetry are given by
 \be
 \label{exteriords2}
 ds^2=-dt^2+\beta^2 dz^2
 +b^2(t)\left[(dr-\epsilon r dz)^2+
 \Sigma^2(r,k) d\varphi^2\right],
 \ee
 where $\beta$ is a constant, $\Sigma$ and $k$ are given by (\ref{sigma}),
 and $\epsilon=0,1$ is such that $\epsilon k=0$.\finn
 \end{theorem}
 {\bf Remark:}
 The line-element for the static region then becomes
 $$
 ds^2=
 -A^2dT^2-2cA^2dTd\tilde z+d\rho^2+C^2 d\tilde{\varphi}^2+
 (\beta^2+\epsilon C^2)d\tilde{z}^2,
 $$
 where $A$ and $C$ are functions of $\rho$ and $c$ is constant.
 The matching hypersurface
 is given by (\ref{surf+}) with
 $\dot f_{z}(\lambda)\eqq
 \epsilon r\dot r b^2/(\beta^2+\epsilon r^2 b^2)$
 and by (\ref{surf-}) with
 $\dot f_{\tilde\varphi}(\lambda)=0$
 and $\dot f_{\tilde z}(\lambda)\eqq c A^2/(\beta^2+\epsilon r^2 b^2)$,
but it is not fully determined in general. The set of equations
to be satisfied are those given in (\ref{hola31}) and (\ref{h13exp}).

 This demonstrates that the possible LRS models (\ref{exteriords2})
 are extremely special. In fact,  
 condition $a(t)=\beta$ on (\ref{compact}) with $n=0$
 poses a strong constraint, whereby
 the timelike surfaces $\Omega$ in (\ref{exteriords2})
 parametrized by $\{\lambda_1,\lambda_2\}$ and
 defined by $\{t=\lambda_1, z=\lambda_2,
 \varphi=\varphi_0, r=r_0 e^{\epsilon \lambda_2}\}$,
 where $\varphi_0$ and $r_0$ are constants,
 have no dependence on time.
 As mentioned in Section \ref{compact-lrs},
 in the $n=0$ cases of (\ref{compact}), and thus in (\ref{exteriords2}),
 the surfaces $\Omega_S$
 spanned by $r$ and $\varphi$ (at constant $t$ and $z$)
 are surfaces of constant curvature,
 since there is a $G_3$ acting multiply transitively,
 which is generated by $\vec\eta_1, \vec\eta_3$ and $\vec\eta_4$
 (\ref{akv}), (\ref{kkk}).
 When $\epsilon=0$,
 the family of surfaces $\Omega$ are just the family of orthogonal
 surfaces to the $\Omega_S$ orbits.

 Another way of looking at this is that
 one of the components of the
 expansion tensor $\theta_{\alpha\beta}$($=\nabla_{(\alpha}u_{\beta)}$)
 of the flow given by $\vec u=\p_t$
 vanishes. % in this case.
 To be more precise, the only non-vanishing
 components of $\theta_{\alpha\beta}$ in the natural orthonormal tetrad
 \bea
 \bm{\theta}_0=dt,~~\bm{\theta}_1=b(dr-\epsilon r dz),
 ~~\bm{\theta}_2=b\Sigma d\varphi,
 ~~\bm{\theta}_3=\beta dz,
 \label{cobasis}
 \eea
 %$\{dt,\beta dz,b(dr-\epsilon r dz),b\Sigma d\varphi\}$
 are
 \be
 \label{tet}
 \theta_{11}=\theta_{22}=b_{,t}/b.
 \ee
 % and $\theta_{33}=0$.
 This is a strong constraint as far as 
 cosmologically interesting models are concerned,
 since there is %then an observer %, given by $\vec u$,
 no expansion along the spacelike direction
 spanned by $\p_z+\epsilon r \p_r$ (which is orthogonal to $\Omega_S$
 iff $\epsilon=0$).

 This result can be seen in two ways: either
 as a consequence of the assumption that the 
 metric $g^-$ is static and cylindrically symmetric,
 or as a consequence of
 the homogeneity in the evolving spacetime,
 which prohibits the norm of
 the Killing vector $\p/\p z$
 to be space dependent.
 %So, it is possible that
 The condition $a(t)=$constant may not be necessary 
 if either the assumption of the cylindrical symmetry on $g^-$
 or the homogeneity of the metric $g^+$ are relaxed,
 although one might still expect strong constraints on $g^+$
 leading to restrictions on the possible matter content there.
 We shall return to these questions in a future publication.

 So far we have not restricted the source fields in the matching spacetimes. We shall now consider
 the particular case of a perfect fluid LRS metric. 
  
 %______________________________________________________________
 \subsection{Perfect-fluid LRS region}
 %______________________________________________________________
 The Einstein tensor for (\ref{exteriords2})
 in the natural orthonormal tetrad (\ref{cobasis})
 %$\bm \theta^\alpha \propto dx^\alpha$
 has the form 
 \begin{eqnarray}
 G_{00}&=&\frac{k+b^2_{,t}}{b^2}-3\frac{\epsilon}{\beta^2}\nonumber\\
 G_{03}&=&-2 b_{,t}\frac{\epsilon}{\beta b}\nonumber\\
 G_{11}&=&G_{22}=-\frac{b_{,tt}}{b}+\frac{\epsilon}{\beta^2}\label{eins}\\
 G_{33}&=&-2\frac{b_{,tt}}{b}-\frac{k+b^2_{,t}}{b}
 +\frac{\epsilon}{\beta^2}.\nonumber
 \end{eqnarray}
The allowed Segr\'e types are $\{1,1(11)\}$, $\{2(11)\}$ together with
 their degeneracies. We are interested in the perfect-fluid type, 
 i.e. $\{1,(111)\}$.
 The first condition for this type of
 source is $(G_{03})^2=(G_{00}+G_{22})(G_{33}-G_{22})$
 (see \cite{MATHO,Vera-thesis}), which can explicitly be cast in the
 form
 \be
 \label{pf}
 \left(b b_{,tt}-b_{,t}^2-k\right)
 \left[\beta^2\left(b b_{,tt}+b_{,t}^2+k\right)+2\epsilon b^2\right]=0,
 \ee
 so that $\rho=G_{22}+G_{00}-G_{33}$ and $p=G_{22}$.
 The vanishing of the 
 %Let us consider the 
 first term in (\ref{pf}) results 
 %If that term vanishes then we have 
 in $G_{00}+G_{22}=-2\epsilon/\beta^2$.
 In order to have a perfect fluid we also need $G_{00}+G_{22}\neq 0$
 to have the same sign as $\rho+p$. As a result,
 we are left with the case $\epsilon=1$,
 which implies $\rho+p<0$.
 In this case both $\rho$ and $p$ are constants
 such that $\rho+3 p=0$.

 Therefore, in order to have a perfect fluid
 satisfying the dominant energy condition (i.e. $\rho+p>0$)
 we can only consider the vanishing
 of the second term in (\ref{pf}).
 We shall consider the cases $\epsilon =1,0$ in turn.
 For the case $\epsilon=1$, equation
 $\beta^2\left(b b_{,tt}+b_{,t}^2\right)+2 b^2=0$ gives $b(t)=
 c_1 \sqrt{\sin(2t/\beta)}$ (after rescaling $t$),
 and hence $\rho=p-8/\beta^2=1/\beta^2(1/\sin^2(2t/\beta)-6)$. The energy
 density changes sign at $\sin(2t/\beta)=1/\sqrt{6}$, and
 thus the weak energy condition cannot be fulfilled over the whole
 %over the
 spacetime.
 %manifold.

 Therefore, we are left with the case $\epsilon=0$; in which case the equation for
 $b(t)$
 becomes 
 %\be
 %\label{eq-B}
 $b^2_{,t}+b_{tt}b+k=0,$
 %\ee
 giving
 \be
 \label{eq-B3}
 b(t)=\sqrt{\alpha t-kt^2},
 \ee
 where $\alpha$ is an arbitrary constant that can be taken to be positive
 without loss of generality.
 This corresponds to a stiff perfect fluid
 given by
 \be
 \label{stiff}
 \rho=p=\frac{\alpha^2}{4 t^2(\alpha-k t)^2},
 \ee
 which ensures that the energy conditions are satisfied.
We recall that this solution can also be interpreted as applying
to the case with a minimally coupled scalar field as the source. 
These results are summarized in the following theorem:
 \begin{theorem}
 \label{bigone}
 The only possible non-static $G_4$ on $S_3$ LRS
 perfect-fluid spacetimes satisfying the dominant energy condition
 that can be matched to an OT cylindrically
 symmetric static metric across a non-space-like hypersurface preserving
 the symmetry are given by
 \be
 \label{exteriorpf}
 ds^2=-dt^2+dz^2
 +(\alpha t-k t^2)\left[dr^2+
 \Sigma(r,k)^2 d\varphi^2\right],
 \ee
 where $\alpha$ is a constant
 and $\Sigma$ is defined as in (\ref{sigma}).
 The equation of state is that of
 a stiff fluid and is given by (\ref{stiff}).
 \end{theorem}

 This amounts to a no--go result, namely 
 that there are no evolving $G_4$ on $S_3$
 LRS perfect-fluid spacetimes with $\rho\neq p$, fulfilling
 the dominant energy condition, that match
 a locally OT cylindrically symmetric static region across
 a non-spacelike matching hypersurface preserving the symmetry.

 So far we have studied the matching between a $G_4$ on $S_3$
 LRS region and an OT
 cylindrically symmetric static region across a non-spacelike
 matching hypersurface preserving the symmetry. This 
 treatment has been local
 and has not dealt with
 matching in the specific context of a particular 
 configuration. Consequently, our results can be used
 in a number of different settings.
 For example, they can be employed to study the generalization of 
 the Einstein-Straus result concerning the embedding of a static
 region in a LRS cosmological model, by taking
 the static part as describing {\em locally} an interior region and the LRS
 part as its exterior.
 But given the interior/exterior duality in the matching procedure 
 they can also be used to study
 the question of the existence
 of an astrophysical evolving object described {\em locally}
 by a LRS metric which is surrounded by an OT cylindrically
 symmetric static background. In this way our results can 
 be used to consider
 generalizations of the Oppenheimer--Snyder \cite{OPSN}
 collapsing model.

 The no-go result above then tells us
 that {\em a $G_4$ on $S_3$ LRS cosmological model cannot contain
 a locally OT cylindrically symmetric static cavity}
 except for the very particular stiff-fluid case mentioned above.
 Theorem \ref{bigone} rules out not only static cosmological strings
 in LRS cosmological backgrounds, but also static cavities
 which are locally cylindrically symmetric, as for instance, bottle/coin-shaped
 objects.
 Furthermore, it implies that {\em no astrophysical
 object described by a non-stiff fluid type
 $G_4$ on $S_3$ LRS metric can be embedded into a
 locally OT cylindrically symmetric static background}.
 Importantly, these results hold 
 %We remark again that all of this is
 irrespective of the matter content in the static part.

 Concerning global 
 %sBut taking into account not only the local problem but also the global
 configurations, as discussed above,
 %as in these last considerations, we can still
 we can go further and apply our results to the case of spatially
 homogeneous non-static exterior spacetimes.
 This follows from the fact
 that the model for a spatially bounded interior
 region whose bounding surface is topologically $S^2$
 preserves the existence of an axis of symmetry
 across this border.
 This implies that the exterior homogeneous part
 has to be locally rotational symmetric, and
 thus admit a further isometry becoming
 a $G_4$ on $S_3$ LRS region.
We shall demonstrate this result in the following section.
 
 %_________________________________________________________
 \section{Bianchi spacetimes: axially symmetric global models}
 \label{axialglobal}
 %__________________________________________________________

 We recall
 that a spacetime admits a cyclical symmetry if
 its metric is invariant under an effective realization of the
 one-dimensional torus
 on the manifold \cite{carter70}.
 Axial symmetry arises when the set of fixed points is non-empty
 (i.e. the generator of the isometry, say $\vec \xi$, vanishes).
 In fact, it has been shown \cite{maseaxconf} that any
 non-empty set $W_2$ of fixed
 points in a four-dimensional spacetime is
 a timelike two-dimensional surface. Furthermore,
 the axial Killing vector field $\vec \xi$ is spacelike
 in a neighbourhood of the axis and satisfies 
 the regularity condition, i.e. the expression (\ref{eq:condregaxis})
for $\vec\xi$ tends to 1 on $W_2$.
 % \be
%  \left.\frac{\nabla_\rho({\vec \xi}^2)\nabla^\rho({\vec \xi}^2)}{4{\vec \xi}^2}
%  \right|_{W_2}\longrightarrow 1.
%  \label{eq:condregaxis}
%  \ee
 Here we shall concentrate on the preservation of
 an axial symmetry across a matching surface.
 The models for the common compact and simply connected astrophysical objects, 
 usually consist of
 an interior region whose spatial boundary is topologically a two-sphere $S^2$.
 The boundary $\sup$
 %(we will use all throughout this paper the minus
 %sign to refer to the interior region and the plus
 %for the exterior), 
 (and thus $\sup^+$ and $\sup^-$)
 is then taken to be non-spacelike and
 homeomorphic to $S^2\times I$, where $I$
 is an open interval of the real line. 
 In other words, it is assumed
 that $\sup$ can be foliated
 by a set of spacelike two-surfaces $S_{\tau}$ homeomorphic
 to $S^2$.
 We refer to \cite{Mars,Mars01} for a detailed general construction.
 The surfaces $S_{\tau}$ are embedded into $\mmm^+$ (respectively $\mmm^-$)
 by the maps $\Phi^+_{\tau}\equiv \Phi^+\circ i_{\tau}$
 (respectively $\Phi^-_{\tau}\equiv \Phi^-\circ i_{\tau}$)
 where
 $i_{\tau}:S_{\tau}\rightarrow \sup$ are the natural inclusion
 for each surface into $\sup$.

 Let us denote by
 $\vec\xi^+$ the generator
 of a spacelike (cyclic) isometry on
 $(\mmm^+,g^+)$.
 Since we wish to preserve this symmetry across $\sigma$,
 there exists a vector field $\vec \gamma$ defined in $\sup$ such that
 $d\Phi^+(\vec \gamma)=\vec\xi^+|_{\sup^+}$.
 That is, the restriction of $\vec\xi^+$
 on $\sup^+$ is tangent to $\sup^+$ everywhere.
 Let us now take $(\mmm^+,g^+)$ to be a spatially homogeneous
 spacetime. We can then
 construct a natural foliation of
 the manifold $\mmm^+$ %(or an open subset)
 by taking the homogeneous spacelike hypersurfaces, say
 $\{t= const.\}$
spanned by the orbits of the simply-transitive $G_3$ on $S_3$
 group of isometries.
 By construction, the restrictions of our Killing vector field
 to the orbits $\vec\xi^+|_{\{t\}}$
 are tangent to these hypersurfaces.
 Since $\sigma^+$ is non-spacelike everywhere,\footnote{This assumption
 can in fact be replaced by a less restrictive one, see \cite{Mars01}.}
 we can now define the following foliation $\{S^+_t\}$ of $\sigma^+$:
 $S^+_{t}\equiv
 \sup^+\cap \{t=const.\}$, where $S^+_{t}$ is taken to be the image
 of $S_{t}$ through $\Phi^+_{t}$. The restriction of
 $\vec\xi^+$ on $S^+_{t}$, $\vec\xi^+|_{S^+_{t}}$, is clearly tangent to
 the surfaces $S^+_{t}$, and therefore
 there is a vector field $\vec\gamma_t$ defined in $S_{t}$ such that
 %$d i_t(\vec \gamma_t)=\vec\gamma|_{S_t}$, and so
 $d\Phi^+_{t}(\vec\gamma_t)=\vec\xi^+|_{S^+_{t}}$.

 We now demand that this foliation is
 such that $S_{t}$ is homeomorphic to $S^2$, that is,
 we take $S_{t}$ to be the $S_{\tau}$ above\footnote{This
 might not be necessary,
 as in most cases one may be able to find a diffeomorphism
 between a previously constructed foliation $S_{\tau}$ and
 the surfaces given by $S_{t}$.}.
 $ $ In the following we shall refer to ``spatially compact'' as
 ``spatially compact according to the homogeneous slicing''.
 There must then exist a point where $\vec\gamma_t$ vanishes \cite{fulton}.
 So, for every $t$ there exists a point
 $w_t\in S_t$ where $\vec\gamma_t=\vec 0$,
 and hence
 \be
 \label{isot}
 \vec\xi^+|_{w^+_t}=\vec 0,
 \ee where
 $w^+_t=\Phi^+_t(w_t)$.
 The existence
 of a fixed point for the cyclic symmetry
 (generated by $\vec\xi^+$)
 ensures the existence of a timelike surface of fixed
 points $W^+_2$. Furthermore, there is a neighbourhood around any point
 in $W^+_2$ where $\vec\xi^+$ is spacelike and vanishes only
 at $W^+_2$ \cite{maseaxconf}.
 This can be used to show that
 the points $w^+_t$, for all $t$, are in $W^+_2$.
Actually, because $\vec\gamma$ generates a cyclic symmetry in $\sigma$,
which is inherited from the embedding (see, for instance, \cite{Vera01}),
the set(s) of fixed points of $\vec\gamma$ must be timelike curves
in $\sigma$,\footnote{This comes from the fact that
$\nabla_a \gamma_b$ is of rank 2. See \cite{Marctesi}.}
defined as $W\equiv \{ w_t; \forall t\in I\}$.
Also, since $d\Phi^+$ is a rank-three map,
$\vec\xi^+|_{\sigma^+}$ can only vanish where $\vec \gamma=0$,
i.e. on the curve $W$,
and thus $W^+_2$ cannot be %entirely
contained in $\sigma^+$.
 It follows that
 $\mmm^+ \backslash \sigma^+$ contains points on the axis of symmetry $W^+_2$
 of the cyclic (in this case axial) symmetry generated by $\vec\xi^+$.
 As a consequence 
 $(\mmm^+,g^+)$ cannot be completely anisotropic and spatially homogeneous,
as otherwise there would not be
 a Killing vector field with zero points;
 $(\mmm^+,g^+)$ must at least admit one isotropy, which
 is generated by the Killing vector field $\vec\xi^+$.

 Furthermore, since
 $d\Phi^-_{t}(\vec\gamma_t)=0$ at $w^-_t=\Phi_{t}^-(w_t)$,
 we will also have a non-empty set of points where
 one generator, say $\vec\xi^-$, of the cyclic symmetry
 we are preserving on $(\mmm^-,g^-)$
 will vanish.
 These points $w^-_t$ are precisely those
 that are to be identified with $w^+_t$.
 The same argument as above can then be used
 for the existence of an axis $W^-_2$
 in $\mmm^- \backslash \sigma^-$.
 We have then shown the following:

 \begin{lemma}
 \label{llrs}
 Let $(\mmm, g)$ be a spacetime resulting from the matching
 of two cyclically symmetric spacetimes $(\mmm^+,g^+)$ and $(\mmm^-,g^-)$
 preserving the symmetry.
 If one part, say $(\mmm^+,g^+)$, is spatially homogeneous and
 either part represents a spatially compact and simply-connected
 region,
 %(according with the homogeneous slicing),
 then both $(\mmm^+,g^+)$ and $(\mmm^-,g^-)$ must be axially symmetric.
 In particular, $(\mmm^+,g^+)$ is locally rotationally symmetric admitting
 a $G_4$ on $S_3$ group of isometries.\fin
 \label{needaxis}
 \end{lemma}
 Note that since this Lemma
 relies on the topology of the matching boundary, the spatially
 homogeneous region ($+$) does
 not necessarily correspond to an exterior region, and 
 therefore
 ($-$) and ($+$) can be interpreted as either interior or exterior.

 Using this lemma %\ref{needaxis}
 we can apply 
 the results given in the previous sections for $G_4$ on $S_3$
 LRS spacetimes to Bianchi spacetimes, once one of the
 regions of the matching represents a bounded object
 without holes.
 Since in this paper we have mainly focused on the
 generalization of the Einstein-Straus model,
 we shall, in the following statements, consider 
 the static region to be the spatially bounded cavity
 surrounded by a homogeneous background.
 $ $From Lemma \ref{llrs} we obtain

 \begin{coro}
 A non-static homogeneous Bianchi II, VIII or IX spacetime cannot
 be matched to a spatially compact and simply connected locally cylindrically
 symmetric static region across a non-space-like hypersurface preserving the
 symmetry.
 \end{coro}
 And similarly, from Theorem \ref{loi} we have:

 \begin{coro}
 \label{loicoro}
 The only possible non-static spatially homogeneous spacetimes
 that can be matched to a spatially compact and simply connected
 locally cylindrically symmetric static region
 across a non-space-like hypersurface preserving
 the symmetry are given by (\ref{exteriords2}) with $k=0,-1$.
 \end{coro}
 The same remarks made about Theorem \ref{loi} regarding the interior
 apply here. Finally, from Theorem \ref{bigone} we obtain

 \begin{coro}
 \label{bigonecoro}
 The only possible non-static spatially homogeneous perfect-fluid 
 spacetimes,
 satisfying the dominant energy condition,
 that can be matched to a spatially compact and simply connected
 locally cylindrically symmetric static region,
 across a non-space-like hypersurface preserving
 the symmetry, is given by (\ref{exteriorpf}).
 The possible Bianchi types of the $G_3$ on $S_3$
 are I, III or VII$_{0}$ and the equation of state is that of a
 stiff fluid.
 \end{coro}
 The last corollary amounts to a no-go result, namely
 that {\em there is no possible evolving perfect-fluid
 Bianchi spacetimes with $\rho\neq p$ satisfying
 the dominant energy condition and containing
 a locally OT cylindrically symmetric static cavity
 preserving the symmetry.}

 The above corollaries have
 focused on the existence of static cavities
 in homogeneous backgrounds.
 Given the interior/exterior 
 duality and taking  
%of the interior and exterior and taking
 into account the remark after Lemma \ref{llrs},
 the same results also apply when the spatially homogeneous
 region is taken to be the bounded region which is embedded
 in a locally OT cylindrically symmetric static background.
 They therefore
%to the the case of the Einstein--Straus model, this now provides 
provide a no--go result concerning the anisotropic
generalisation of the Oppenheimer--Snyder model.

 %____________________________________
 \section{Conclusion}
 \label{conclusion}
 %______________________________________
 We have studied
 %, in particular,
 a generalisation of the Einstein-Straus model, by considering a 
 locally cylindrically symmetric static cavity
 embedded in an expanding
 LRS region. 
 We have derived the matching conditions
 for such space-times and have found that
 they 
 impose strong 
 constraints on the LRS metrics, by implying that $a_{,t}=0$ and $n=0$.
 The former implies that no dynamical evolution is allowed 
 along a spacelike direction as seen by the observer $\p_t$.
 This direction is orthogonal to the orbits of the subgroup
 $G_3$ on $S_2$ of the $LRS$ when $\epsilon=0$.
Condition $n=0$ implies
that it is impossible 
to have an exterior metric of Bianchi types $II,VII$ or $IX$.
Our main result in this connection, expressed in
Theorem \ref{loi} and Corollary \ref{loicoro},
is that the exteriors can only take very particular forms
within the Bianchi types $I,III,V,VII_0,VII_h$ or
Kantowski-Sachs metrics.
These results make no reference to the matter content and 
are therefore, in 
this sense, completely general.

 To study the effects of including 
 matter contents,
 we also considered perfect fluid
sources for the metrics allowed in Theorem \ref{loi},
  and found 
 that such embeddings are only possible 
 when the matter content is a stiff fluid.
 This is our second main result, which is summarized in
 Theorem \ref{bigone} (and Corollary \ref{bigonecoro}) .

 We have also proved that if the non-static spacetime is assumed to be
 spatially homogeneous (not necessarily LRS) and if the static
 spacetime represents a spatially compact and simply-connected region,
 then in order to perform the matching preserving the cyclic symmetry
 the non-static part must be LRS. As a consequence, we were able to
 reformulate our results with the weaker assumption of homogeneity,
 instead of local rotational symmetry.
  
 Given that deviations from isotropy and sphericity
 are expected to be present in the universe, these results are
 of potential interest, since
 they make it impossible to embed locally OT cylindrical static
 objects (which are compact and simply connected) in
 %anisotropic
 homogeneous universes
 and at the same time have a reasonable 
 exterior cosmological evolution. 

 Due to the interior/exterior duality,
 our results also apply to the cases of 
 bounded objects
 described by spatially homogeneous metrics embedded
 in locally cylindrically symmetric static backgrounds.
 In particular, this would have the interesting consequence that
 the Oppenheimer-Snyder
 model for collapse cannot be generalised in this way.

 Finally it would be of interest to study
 the inhomogeneous generalisations of our results.
 We hope to return to this question in a
 future work.

%%%%%%%%%%%%%%%%%%%%%%%%%%%%%%%%%%%
\vspace*{-2pt}
\section*{Acknowledgments}
We thank M. MacCallum, M. Mars and J.M.M. Senovilla for
very fruitful discussions and interesting comments. 
FCM thanks CMAT, Universidade do Minho for 
support and FCT (Portugal) for grant PRAXIS XXI BD/16012/98.
RV thanks EPSRC for grant MTH 03 R AJC6.
%%%%%%%%%%%%%%%%%%%%%%%%%%%%%%%%%%%

 %_____________________________________________________________________________
 
 %_____________________________________________________________________________
  
 %_____________________________________________________________________________
 \end{document}